\documentstyle[osa,manuscript,psfig]{revtex} 

\begin{document}                                                          
\titlepage

\title{ 
 What can the final state particles in diffractive lepton-nucleon 
scattering tell us?}
\author { Meng Ta-chung and Zhang Yang\\
 {\it Institut f\"ur theoretische Physik, FU Berlin,
    Arnimallee 14, 14195 Berlin, Germany}}
 \maketitle    
                                                       
\begin{abstract}

It is pointed out that "the colorless 
objects(s)" in diffractive lepton-nucleon scattering in the small-$x_B$
region can be probed
by examing the final state particles in such processes:
(A) Perform multiplicity and/or energy factorial moment analyses, and check
especially the $x_B$- and $Q^2$-dependence of the intermittency-indices.
(B) Examine the diffractive structure function
$F_2^{D(3)}(\beta,Q^2;x_P)$ as function of $x_P$
for different
$\beta$ and $Q^2$. It is shown that such analyses and measurements
can yield useful information on the hadronization process and on
the intrinsic properties of the colorless objects. In particular,
the observation of power-law distributions of
$F_2^{D(3)}(\beta,Q^2;x_P)$ with respect to $x_P$ and with respect to 
$x_B$ --- independent of
$\beta$ and $Q^2$-values may be considered as strong indication for the 
existence of self-organized criticality in gluon-cluster formation
processes.
\end{abstract}

\newpage
 
The observation of large rapidity gap events in deep-inelastic lepton-nucleon 
scattering experiments performed in the small $x_B$ region\cite{m1}
shows
that the virtual photons in such processes may encounter 
colorless objects originating from the nucleon. What are such
objects? 
While the existing data\cite{m1} can be reasonably well reproduced by 
Regge-pole models\cite{m2}, it is not known whether/how
the properties and the formation process(es)
 of such objects can be directly 
probed------
in a model-independent manner. Furthermore, deep-inelastic lepton-nucleon
scattering experiments
and empirical analyses in this kinematical region\cite{m1,m3}
 show that the
gluon-density  in the nucleon is 
much higher than those for quarks/antiquarks, and it is 
increasing with decreasing $x_B$\cite{m3}. Are the 
striking phenomena
observed in the small $x_B$ region related 
to one another? In particular, 
can the above-mentioned colorless objects 
be virtual states formed by interacting "soft" gluons
which exist in abundance in this kinematical region?\cite{m4}
Can the observed
final state particles tell us something about the space-time
properties of the colorless object? Is it possible to
extract such information
 directly from experiments?

In  the present note we show that 
these questions can be answered in the affirmative.
Because of the complexity of such process,
we propose to use 
concepts and methods of statistical physics and study distributions,
average values, and
fluctuations
of the relevant observables
which describe the final-state particles
in the appropriate kinematical regions.
To be more precise, we suggest the
following: (A) Perform multiplicity
(and/or energy) factorial moment analyses to obtain the indices
of intermittency for different $Q^2$- and $x_B$-values. 
(B) Examine the diffractive structure function
$F_2^{D(3)}(\beta,Q^2;x_P)$ as function of $x_P$ for different values of
$\beta$ and $Q^2$. 

In the analyses mentioned in (A), we do the following:
(i) Consider event-by-event, at given values of $Q^2$ and $x_B$,
 the data for diffractive lepton-nucleon scattering in which 
the momentum-fraction (often denoted by $x_P$) of the exchanged colorless 
object is known.
The latter quantity can be measured by detecting the final state of
the outgoing proton  (or nucleon-resonances which have
 similar quantum numbers as the proton)
or by determining it approximately by measuring $Q^2$ and $M_x^2$
in such processes.
(ii) Focus attention on the multiplicities
of the 
charged hadrons produced in the collision between the virtual photon
and the colorless object. That is, look at the multiplicities
of the hadronic system which yields the "$M_x$-spectrum", and use them
to calculate the normalized factorial
moments (FM) $F_q$ as functions of the resolution of the phase
space.
To be more precise, consider the multiplicities of the hadrons in
a given phase space interval characterized
by $\Delta\eta,\Delta p_\bot,\Delta\phi $, (pseudorapidity,
transverse momentum and azimuthal angle of the observed particle).
Divide the above-mentioned phase space into $M$ subintervals
and calculate\cite{z1}
\begin{equation}
\label{e1}
F_q={1\over M}\sum^M_{m=1}{\langle n_m(n_m-1)\cdots (n_m-q+1)\rangle\over
\langle n_m\rangle^q},
\end{equation}
where $n_m$ is the multiplicity in the $m$th subinterval, and $\langle\cdots
\rangle$ means taking the average over events. Use the obtained results to
calculate the intermittency
index $\varphi_q$, from the
$F_q$ vs M plots.
(iii) Repeat the procedure mentioned in i and ii for a fixed 
$x_P$, 
a fixed $Q^2$, but for different $x_B$ values.
(iv) Repeat the procedure mentioned in i and ii 
for a fixed (or integrated) $x_P$, 
a fixed $x_B$, but for
different $Q^2$ values.

The proposal mentioned in (A)
 is based on the following reasonings:

Factorial moment analyses can be used to extract information on
dynamical processes in a model-independent way.
In particular, if dynamical fluctuations have a typical size (e.g.
$\delta \eta_0$ in one-dimensional pseudorapidity phase space
due to resonance
decays which have a width $\delta \eta_0=1$) the factorial moments rise
with decreasing bin-size in pseudorapidity-interval $\Delta\eta /M$
 as long as $\Delta \eta /M >\delta \eta_0$ and
saturate to some constant value for $\Delta \eta /M<\delta\eta_0$.
But, if self-similar fluctuations exist at all scales of $\Delta\eta /M$,
the factorial moments are expected to follow a power-law
$F_q\sim M^{\varphi_q}$ in the limit $M\to \infty$
which manifests itself in the corresponding $\ln F_q$ vs $\ln M$
plot as a linear rise of the slope. Such a behaviour is known\cite{z1,z2} 
as intermittency, and its strength is characterized by the slope parameter
$\varphi_q$ which is also called the anomalous scaling index.

The $x_B$-dependence of such $F_q$ vs $M$ plots, and in particular
that of
 $\varphi_q(Q^2, x_B; x_P)$ for fixed $Q^2$ (and $x_P$), 
plays a distinguished role in such studies. This is because, viewed
from a fast moving reference frame, for example the lepton-nucleon
center-of-mass frame,
 where the nucleon's momentum $\vec P$ is large in high-energy
collisions, we see 
in a diffractive
scattering event\cite{m1}
the following: 
The virtual photon $\gamma^*$ originating from the incident lepton is
absorbed by the nucleon. Precisely speaking,
its energy-momentum $q\equiv (q^0,
\vec q)$ is absorbed by a virtual colorless object of the nucleon. 
 The time interval $\tau_{\rm int}$ in which the absorption
process takes place 
(it is known as the lepton-nucleon 
interaction/collision time)
can be estimated by making use of
the uncertainty principle. In fact, by calculating 
$1/q^0$ in this reference
frame we obtain:
\begin{equation}
\label{e2}
\tau_{\rm int} = {4|\vec P| \over Q^2} ~ {x_B\over 1-x_B}.
\end{equation}
This means, for given $\vec P$ and $Q^2\equiv -q^2$, 
$\tau_{\rm int}$ is directly proportional to
$x_B$ for $x_B<<1$. In other words, {\it $x_B$ is a measure of
the time-interval in which the absorption of $\gamma^*$
by the space-like virtual colorless
object
takes place.}
Hence, by studying the $x_B$-dependence of the intermittency
index $\varphi
_q(x_B,Q^2; x_P)$, we  are 
not merely probing the 
statistical and dynamical fluctuations of the collision process between 
$\gamma^*$ and 
the colorless object
which we hereafter call $c_0^*$.
Since this
process of hadronization of the virtual colorless object $c_0^*$
is initiated by
the interaction with $\gamma^*$,
we are also examing
whether (if yes, how) the hadronization process 
changes with the interaction time $\tau_{\rm int}$. 
This question is of considerable interest, because 
a virtual photon $\gamma^*$
can (logically) only be absorbed by virtual systems ($c_0^*$'s) whose
lifetimes ($\tau_c^*$'s) are longer than $\tau_{\rm int}$ (i.e. $\tau_{\rm
int}\le
\tau_c$). 
That is, the
average lifetime $\langle
\tau_c\rangle$ of the $c_0^*$'s, which can absorb a $\gamma^*$
associated with interaction-time
$\tau_{\rm int}$,
 is a function of the number of $c_0^*$'s which satisfy the condition
$\tau_{\rm int}\le\tau_c$.
Hence, from the 
$x_B$-dependence of the scaling behaviour of FM's,
in particular from that of the corresponding $\varphi_q$'s,
we can find out whether/how
the process of hadronization of a $c_0^*$ depends on its average
lifetime $\langle\tau_c\rangle$ of the $c_0^*$'s. This means,
by measuring the above-mentioned
$x_B$-dependence we can in principle obtain information on 
the number-distribution 
of such colorless objects as a function of their lifetimes. 
It should be mentioned at this place however that, in many cases 
(an example will be given below) multiplicities depend first of all on the
invariant mass of the produced particles. Hence, special care has to be 
taken to conjecture such associations.
See in this connection, also the method proposed in (B) below.

The $Q^2$-dependence of
the scaling behaviour of the FM's, in particular
that of $\varphi_q(Q^2,x_B; x_P)$,
 is also of considerable interest.
This is because, in photon-proton scattering experiments, 
not only those with real $(Q^2=0)$ photons but also
those with
space-like $(Q^2>0)$ photons where $Q^2$ is not 
too large ($\le 1 {\rm GeV}^2/c^2$,
 say)
have very much in common with hadron-hadron collisions.
Having in mind that the index of intermittency
for hadron-hadron scattering is smaller than that for 
electron-positron 
annihilation processes\cite{z2}, 
we are led to the following questions: Do we expect
to see a stronger $Q^2$-dependence when we increase $Q^2$ from zero to 
10 or 100 $GeV^2/c^2$, say? Is this also a way to see whether space-like
photons at large $Q^2$ "behave like hadrons" in such interactions?

While waiting for data to perform the above-mentioned analyes, one
may want to use some existing phenomenological models, for example
JETSET\cite{z4} to generate such "data",
and use them to obtain 
 $F_q$ vs $M$ plots,
and to obtain the $x_B$- as well as the 
$Q^2$-dependence of the intermittency indices for fixed
$Q^2$ and $x_B$ respectively.
The results of such a calculation are predictions based on the assumption
that JETSET\cite{z4} is applicable to diffractive lepton-nucleon scattering.
These prediction can be tested
when the data are available. 
By carrying out such model-calculations, we convinced ourselves that the
proposed method works well.
The details on  this and on other model-calculations 
will be discussed  elsewhere\cite{z3}.

Furthermore, having
in mind that jets have been observed (see e.g. 
 Ref.\cite{m1} and the papers cited there.) in diffractive
electron-proton scattering processes, and experimentally 
calorimeters have been used to measure the collision events in general,
and to study jet-events in particular,
we are naturally led to ask:
Can we, instead of using the distribution of 
multiplicities in phase space, 
also use the distributions of transverse-energies 
to probe the existence of dynamical fluctuations
in such intermittency-analyses?
Here, the transverse-energy $E_\bot$ is measured 
on an event-by-event bases
with respect to
the axis of the virtual photon.
According to the most recent experimental knowledge
(See e.g. the review paper
given in Ref.1), we expect to see that the distributions
of $E_\bot$ in phase space
in such collision events are symmetric with respect to this axis,
and symmetric with respect to the origion of the c.m.s. frame of the colliding
objects----namely $\gamma^*$ (the virtual photon) and $c_0^*$
(the virtual colorless object). As a first step, 
we generalized in a straightforward manner the usual procedure\cite{z1,z2}
by introducing an energy-unit $\varepsilon$
and write the "energy factorial moment" $F_q^{(E)}$ as 
$\langle E_\bot (E_\bot -\varepsilon)\cdots [E_\bot -(q-1)\varepsilon ]
\rangle/\langle E_\bot\rangle^q$. It is clear that
$E_\bot/\varepsilon$ can be considered as integers, provided that
$\varepsilon$ is sufficiently small. 
Under this condition, it is clear that statistical fluctuations
in $F_q^{(E)}$ can be canceled out in the same way as that in
$F_q$ defined in Eq.(~\ref{e1}). But, this means, there is 
a dependence on 
an arbitrary parameter $\varepsilon$, when we use $F_q^{(E)}$!
Is there a way to get rid of this kind of arbitrariness in the
practice?

In order to answer this question, 
let us consider a collision
event, in which the total transverse energy is $E_\bot$(total), and
the ratio between the arbitrarily chosen $\varepsilon$ and
$E_\bot$(total) is $\lambda$ i.e. $\lambda\equiv \varepsilon/E_\bot$(total).
We generate events under the assumption that 
$E_\bot /\varepsilon$ obeys the Bernoulli distribution
for different sizes of subintervals, that is,
when the above-mentioned number M changes. It is clear
that the slope in the double logarithmic
$F_q^{(E)}$ vs M plot has to be flat.
Next, we introduce
the moment 
$R_q^{(E)}\equiv\langle E^q_\bot\rangle/\langle E_\bot\rangle^q$
and calculate this quantity
for different choices of $\lambda$.
The corresponding $R_2^{(E)}$ vs M plots
is shown in Fig.1.
Here we see that $R_2^{(E)}$
can be considered a good approximation for
$F_2^{(E)}$, when $\varepsilon$ is of the order of $10^{-3}$ of
the total $E_\bot$ in the event under consideration. 

Let
us recall that, by studying $F_q$ we are examing the anomalous
scaling behaviour\cite{z1} of the 
probability-moments for multiplicies $C_q\equiv 1/M\sum^M_{m=1}
\langle p_m^q\rangle/\langle p_m\rangle^q$;
by measuring $R_q^{(E)}$, we are looking at that of the corresponding
probability-moments for transverse energies. Since the
$R_q^{(E)}$ vs M plots for transverse energies distributions are in
general different from the $F_q$ vs M plots for multiplicities,
comparisons between the two kinds of plots, together with
the $x_B$- and $Q^2$-dependence of the
corresponding
 intermittency indices,
will raise a number of questions, in particular the following: Is
the multiplicity distribution
or the transverse-energy distribution
more relevant for
studying dynamical fluctuations?
It seems that much work is still needed to answer such
questions.

As we have mentioned in the introduction, and have shown in the part (A)
of this paper, our goal is to find out whether/how {\it
model-independent information on the colorless objects can be obtained 
by examing the final state particles in diffractive lepton-nucleon
scattering processes}. For this purpose, we now consider the measurements
mentioned in (B), and discuss the notion as well as the consequencies
of "self-organized criticality". We 
recall that, in
a series of papers, 
P. Bak, C. Tang and
K. Wiesenfeld \cite{z5}(hereafter refered as BTW) 
have shown that certain dissipative dynamical
systems with extended degrees of freedoms can evolve towards
a self-organized 
 critical state which give rise to 
spatial and temporal power-law scaling behaviour.
The spatial scaling leads to self-similar fractal structure. 
The temporal scaling manifests itself as flicker noise.
Having in mind that the density of soft-gluons is high in the 
small $x_B$ region\cite{m3}, and that in an extended system (in space-time)
such gluons may interact with one 
another in accordance with QCD, the facts found by BTW leads us to the
question:
Can systems of gluons of the target proton evolve into 
self-organized critical states? Knowing (at the present time) very little
about the dynamics of soft-gluons on the one hand, and even less
about the general 
dynamical origin of
self-organized criticality on
the other, we are not in a position to answer this question {\it
theoretically}. But, with the help of our colleagues at HERA, 
there seems to be a chance
to answer it {\it experimentally}. 
To be more specific, let us rephrase
 the question 
as follows: If 
the interacting soft-gluons can indeed form such critical states,
where the interaction with an additional gluon
can be considered as  a local perturbation,
what kind of signals do we
expect to see experimentally?

According to the known characteristics of
self-organized states of BTW\cite{z5},
the structures of such states 
are barely stable; and 
a local perturbation
of a critical state 
 can grow over all
length scales, leading to anything from a "shift" of a single
unit to an avalanche, the size of which can be as large as the entire
system. The distribution
of the transported physical quantities, "the
dissipated energy",
is a measurable quantity; and it has been
shown by BTW
(in two and three dimensional cases) that the distribution of this
energy (also called "the size of cluster" by BTW), and
the distribution of the time-interval in which the formation process
of such a cluster takes place (called by BTW
 "the lifetime of the cluster")
always obey {\it power-law}. 

In connection with the expected experimental signals, 
let us recall the following:
Diffractive lepton-proton scattering
can be envisaged as the collision between the virtual photon 
$\gamma^*$ and the virtual colorless object $c_0^*$ which carries a
fraction $x_P$ of the energy of the incident proton (in, e.g.
the above-mentioned electron-proton c.m. frame).
The "diffractive structure function"
 $F_2^{D(3)}(\beta, Q^2;x_P)$ at given
$\beta$- and $Q^2$-values can be considered 
as the probability for $\gamma^*$
(characterized by $Q^2$ and $\beta$) to encounter the $c_0^*$.
Hence, $F_2^{D(3)}$ is simply
the probability for the above-mentioned $\gamma^*$ to "see"
a $c_0^*$ which carries the energy fraction
$x_P$ (in other words,
"the $x_P$-distribution
of the $c_0^*$ in the proton" if it is independent of
$\beta$ and $Q^2$).
This means, for given constant values of $\beta$ and $Q^2$,
we may check whether the $c_0^*$'s can be identified with
BTW-clusters in the following way. Since in that case: (a)
the interacting soft-gluons can evolve
into a state with no characteristic time or length scale, (b) the
gluon-clusters, $c_0^*$'s, are the results of
perturbation of a self-organized critical state
caused by soft-gluon interaction, and (c) "the distributions of the
dissipative energy" (of BTW) are nothing else but the distributions of
energies of the gluon-clusters,
we should see that $F_2^{D(3)}(\beta,Q^2;x_P)$ obeys a power-law
$x_P^{-\alpha}$ with $\alpha\approx 1$, for given constant values of $\beta$
and $Q^2$. Furthermore, keeping
in mind, (as we have already discussed in connection 
with proposal A) that
$x_B$ is a measure of the average lifetime of gluon-clusters;
and that the variables $x_B$, $\beta$ and $x_P$ are related to one another by 
$x_B=\beta x_P$, 
we also should see that
 $\beta^{-1}F_2^{D(3)}(\beta,Q^2;x_P)$ show a power-law behaviour with
respect to $x_B$ (Note that
$\beta$ can be considered as the fraction of the 4-momenta of $c_0^*$
carried by the struck charged constituent interacting with $\gamma^*$).
Let us now look at the experimental data\cite{m1} and examine first
the $x_P$-dependence of
$F_2^{D(3))}(\beta,Q^2;x_P)$ at given values of $\beta$ and $Q^2$.
As we can see in Fig.2 ( Note that the corresponding plots for
$\beta^{-1}F_2^{D(3)}$ vs $x_B\equiv\beta x_P$ have exactly the same form
for given $\beta$!), the existing
data\cite{m1} indeed
show the expected characteristic spatial and temporal scaling
behaviours. These results may
 be considered as strong indication that self-organized
criticality plays a dominating role in the formation of gluon-clusters,
the color-singlet ones of which are the objects we are dealing with
in diffractive lepton-uncleon scattering processes.
Further implications of this observation will be discussed in more detail
elsewhere\cite{z3}.


We  thank C. Boros, D. H. E. Gross,
 Z. Liang, R. Rittel, K. D. Schotte and K. Tabelow
for helpful discussions, 
and FNK der FU-Berlin
for financial support. One of us (Y. Zhang) also thanks Alexander 
von Humboldt  Stiftung for the fellowship granted to him, and the
members of
Institut f\"{u}r Theoretische Physik der FU Berlin for
their warm hospitality.


\noindent
{\large\bf Figure captions}

\vskip 1cm
\noindent
Fig.1, $\ln R_2^E=\ln (\langle E_\bot^2\rangle /\langle E_\bot\rangle^2)$
as functions of $\ln M=\ln (\Delta /\delta )$,  
when the transverse energy $E_\bot$ in units of $\varepsilon$
in subinterval 
$\delta$ is stochastically
produced according to Bernoulli distribution. 
Here, $\lambda =\varepsilon /E_\bot({\rm total})$.

\vskip 1cm
\noindent
 Fig.2, The dependence of $F_2^{D(3)}(\beta,Q^2;x_P)$ upon $x_P$
at given values of $\beta$ and $Q^2$. The data are taken from Ref.\cite{m1}
(The data at lower $Q^2$ are omitted because of lack of space. They
also fall on one straight line in such log-log plots consistent with the data at
large $Q^2$ values). See text for further details.

\begin{thebibliography}{99}

\bibitem{m1}
M. Derrick {\it et al.} ZEUS Collaboration, Phys. Lett. {\bf B315},
481(1993); Phys. Lett. {\bf B332}, 228(1994);
Z. Phys. {\bf C68}(1995)569;
Z. Phys. {\bf C70}(1996)391;
T. Ahmed {\it et al.} H1 Collaboration, Nucl. Phys. {\bf B429}, 477(1994);
Nucl. Phys. {\bf B439}(1995)471;
Phys. Lett. {\bf B348}, 681(1995);
For a recent review, see e.g. J. P. Philips (H1 Collaboration),
Talk given at the 28th Int. Conf. on High-Energy Physics, July 1996,
Warsaw, Poland; and the papers cited therein.

\bibitem{m2}
See e.g. Ref. 1 and the paper cited therein.

\bibitem{m3}
See e.g. S. Aid {\it et al.} H1 Collaboration, Phys. Lett. {\bf B354},
494(1995) and M. Derrick {\it et al.} ZEUS Collaboration, 
Phys. Lett. {\bf B345}, 576(1995).

\bibitem{m4}
F. E. Low, Phys. Rev. {\bf D12}, 163(1975); S. Nussinov, Phys. Rev. Lett.
{\bf 34}, 1286(1975) and Phys. Rev. {\bf D14}, 246(1976);
C. Boros, Liang Zuo-tang and Meng Ta-chung, Phys. Rev. {\bf D54}, 6658(1996).

\bibitem{z1}
  A. Bialas and K. Peschanski, Nucl. Phys. {\bf B273}, 703(1986);
      {\bf B308}, 857(1988).

\bibitem{z2}
For recent review articles, 
see e.g.,
N. Schmitz, in: Proc. XXI Int. Symp. on Multiparticle Dynamics, Wuhan,
1991, eds. Y. Wu and L. Liu (World Scientific, Singapore, 1992) P.377;
and E.A. De Wolf, I.M. Dremin and W. Kittel,
Phys. Rep. {\bf 270}, 1(1996).

\bibitem{z4}
T. Sjostrand, Comput. Phys. Commun. {\bf 39}, 347(1986).

\bibitem{z3}
C. Boros, Meng Ta-chung, R. Rittel and Zhang Yang (in preparation)

\bibitem{z5}
P. Bak, C. Tang and K. Wiesenfeld, Phys. Rev. Lett. {\bf 59}, 381(1987);
Phys. Rev. {\bf A38}, 364(1988). 


\end{thebibliography}
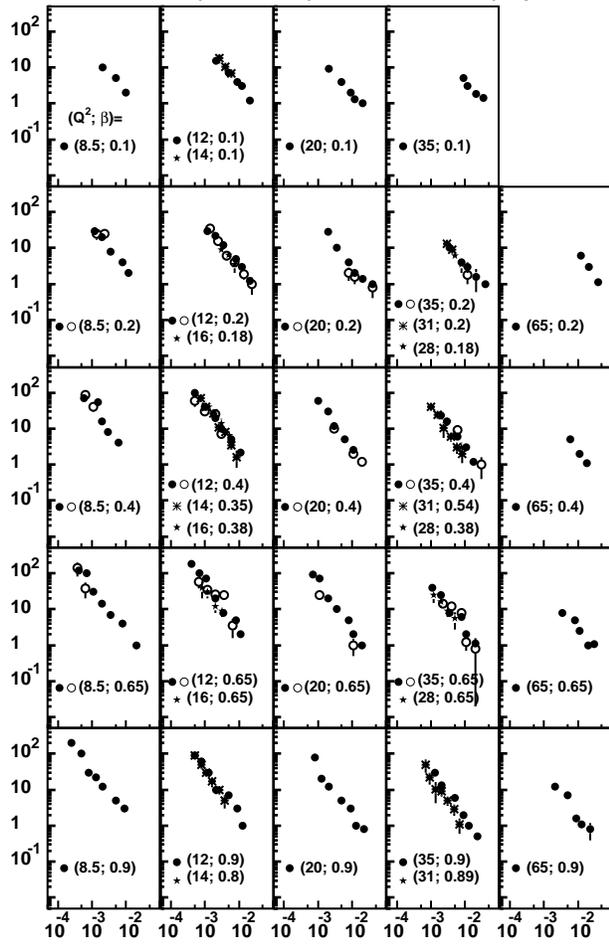
\end{document}